\documentclass[aps,prd,onecolumn,groupedaddress,showpacs]{revtex4}

\usepackage{natbib}
\usepackage[dvips]{graphicx,color}

\def\e{\mathrm{e}}

\begin{document}

\title{Probing the central engine of long gamma-ray bursts and 
hypernovae with gravitational waves and neutrinos}

\author{Yudai Suwa} \email[]{suwa@utap.phys.s.u-tokyo.ac.jp}
\affiliation{ Department of Physics, School of Science, The University
  of Tokyo, Hongo 7-3-1, Bunkyo-ku, Tokyo 113-0033, Japan }
\affiliation{ Max-Planck-Institut f\"ur Astrophysik,
  Karl-Schwarzshild-Str. 1, D-85741 Garching, Germany } 

\author{ Kohta Murase }\email[]{kmurase@yukawa.kyoto-u.ac.jp}
\affiliation{Yukawa Institute for Theoretical Physics, Kyoto
  University, Oiwake-cho, Kitashirakawa, Sakyo-ku, Kyoto, 606-8502,
  Japan }

\preprint{UTAP-614}

\date{\today}

\begin{abstract}
    There are the two common candidates as the viable energy source
    for the central engine of long gamma-ray bursts (GRBs) and
    hypernovae (HNe), neutrino annihilation and magnetic fields.  We
    investigate gravitational wave (GW) emission accompanied by these
    two mechanisms. Especially, we focus on GW signals produced by
    neutrinos from a hyper-accreting disk around a massive black hole.
    We show that neutrino-induced GWs are detectable for $\sim$1 Mpc
    events by LISA and $\sim$ 100 Mpc by DECIGO/BBO, if the central
    engine is powered by neutrinos.  The GW signals depend on the
    viewing angle and they are anti-correlated with neutrino
    ones. But, simultaneous neutrino detections are also expected, and
    helpful for diagnosing the explosion mechanism when later
    electromagnetic observations enable us to identify the source.  GW
    and neutrino observations are potentially useful for probing
    choked jets that do not produce prompt emission, as well as
    successful jets. Even in non-detection cases, observations of GWs
    and neutrinos could lead to profitable implications for the
    central engine of GRBs and HNe.
\end{abstract}

\maketitle

\section{Introduction}

Gamma-ray bursts (GRBs) are the most luminous explosions in the
universe. The production of GRBs is believed to require that only the
small amount of matter is accelerated up to ultrarelativistic speeds
and collimated as a jet \citep{pira05,mesz06}.  The duration of GRBs
ranges from $\sim {10}^{-3}$ s to $\sim {10}^{3}$ s, with a roughly
bimodal distribution of long GRBs of $T \gtrsim 2$ s and short GRBs of
$T \lesssim 2$ s.  The geometrically corrected gamma-ray energy of
long GRBs is typically $E_{\gamma} \sim {10}^{50-52}$ ergs
\citep{frai01,bloo03} (see also \cite{ghir04}), which is much smaller
than the isotropic gamma-ray energy $E_{\gamma}^{\rm iso} \sim
{10}^{52-54}$ ergs.  The requirements for the energy budget and time
scales suggest that long GRBs involve the formation of a black hole
(BH) (e.g., \cite{woos93}) or magnetars via a catastrophic stellar
collapse event.  The discovery of supernovae (SNe) associated with
GRBs brought us the more direct evidence that GRBs result from a small
fraction of massive stars that undergo a catastrophic energy release
event towards the end of their evolution \cite{hjor03,stan03}.
Interestingly, some of the SNe associated with GRBs (e.g., GRB 980425,
GRB 060218) showed evidence for broad lines indicating high-velocity
ejecta with inferred explosion energy of $E_{\rm kin} \sim 10^{52}$
ergs (e.g., \cite{iwam98}), and those energetic SNe are often called
hypernovae (HNe).  The {\it central engine} of both GRBs and HNe has
to supply such an enormous explosion energy (for reviews, e.g.,
\cite{woos06}).

For the class of long GRBs, the commonly discussed candidates are
massive stars whose core collapses to a black hole, either directly or
after a brief accretion episode, possibly in the course of merging
with a companion.  This {\it collapsar} scenario is one of the most
widely believed scenarios to explain the huge release of energy in
GRBs and HNe \citep{woos93,pacz98,macf99}.  In this scenario, the
collapsed iron core of a massive star forms a temporary disk around a
few solar mass BH and accretes at a high rate ($\sim 0.1-10~M_\odot$
s$^{-1}$), which is believed to produce a powerful jet leading to a
GRB. The duration of the burst in this scenario can be related to the
fall-back time of matter to form a disk or the accretion time of the
disk. Possibly, HNe that are brighten by nickel could also be
explained by a disk wind, which is subrelativistic with a speed
comparable to the escape velocity of the inner disk \cite{macf99}.

Provided a disk and BH form, the greatest uncertainty in the collapsar
scenario is the mechanism for converting the disk binding energy or BH
rotation energy into directed relativistic outflows.  So far, two
general mechanisms have been proposed: neutrino annihilation and
magnetohydrodynamical (MHD) mechanisms in various kinds. In the former
case, neutrino pairs are generated in the hot disk and impact one
another with the largest angles along the rotational axis and deposit
some fraction of the accretion energy
\citep{poph99,dima02,kohr02,kohr05,chen07,kawa07,birk07}. It should be
noted that the efficiency of energy deposition is no greater than
$\sim$ 1\% of the total neutrino emission \citep{ruff97}, so that the
required neutrino energy is $E_{\nu} \sim {10}^{53-54}$ ergs in order
to achieve the jet energy of $E_{j} \sim {10}^{52}$ ergs (which should
be larger than the gamma-ray radiation energy).  One important
property of this dense, hot accretion flow is that the density is so
high that cooling of the flow is dominated by neutrinos. This
accretion flow is called neutrino-dominated accretion flow
(NDAF). Although neutrinos can escape the flow more easily than
photons, they can also be trapped in the flow and fail to escape
before being swallowed by the black hole when the accretion rate
becomes very high \cite{poph99,dima02,kohr02}.
The other possible mechanism is a MHD-driven explosion, where two
options exist for the energy source: the binding energy of the
accretion disk \citep{prog03b,naga07} and the rotational energy of the
BH through the Blandford \& Znajek mechanism
\citep{blan77,mizu04b,mcki04,bark08}.  In the former case, the
magnetic fields are amplified up to $\sim 10^{15-16}$ G by rotation
via some disk instabilities, leading to a strong outflow.  In the
latter case, the BH and accretion disk are connected with magnetic
field lines and the rotational energy of the BH is extracted via these
magnetic field lines. In some versions of the MHD scenario, the
extracted energy propagates as the Poynting flux and is probably
converted into a matter-dominated form at later stages, so that the
efficiency of energy deposition is expected to be higher than that of
the neutrino-annihilation mechanism.

Although the various possibilities including the fast rotating
magnetar scenario \cite{usov92,thom94,komi07,dess08,taki09} as well as
the collapsar scenario have been proposed, there has been no direct
evidence for the central engine of GRBs.  In this paper, we consider
gravitational waves (GWs) as a probe of the jet-producing mechanism of
GRBs (see \cite{sago04,hira05} for GWs from jet itself). The
difficulty in probing the central engine comes from the fact that most
observed properties of GRBs come from electromagnetic signals produced
in regions far away from the central engine. On the other hand, GWs
and neutrinos can reach us without losing information of physical
conditions near the central engine.  Especially, we focus on the GW
emission generated by anisotropic neutrino emission, and estimate its
amplitude which were investigated in the context of massive-star
collapse \citep{burr96,muel97,muel04,ott06,kota07,suwa07b} (see also,
\cite{kotarev,ott09}).  The amplitude of this GW emission is governed
by the radiated neutrino energy that has to be large enough for the
neutrino-annihilation mechanism to work as the central engine of GRB
jets. As a result, we can show that neutrino-induced GWs are
detectable for $\sim$1 Mpc events by LISA and $\sim$ 100 Mpc by
DECIGO, if the central engine is powered by neutrinos.  The coincident
neutrino signals can also be detected by using the Super-Kamiokande
detector or recently proposed Mton neutrino detectors such as
Hyper-Kamiokande.  Possible successful detections combined with later
electromagnetic observations can give us important clues to the
central engine of long GRBs.

This paper is organized as follows. 
In \S II, we describe the formulation used in this work. The necessary
expressions of the GW amplitude is presented. In \S III, we give
spectra of GWs from NDAF and apply the derived formula to a burst or
multiple bursts representing GRBs. We also discuss detectability of
GWs and whether we can identify a GW event as the formation of NDAF
making a GRB and/or HN. In \S IV, the gravitational wave background is
also discussed.

\section{Gravitational waves from NDAF}
Here we consider the gravitational wave emission from anisotropic
neutrino radiation, which is proposed by Epstein \cite{epst78}.
Following the formulation developed by previous works
\citep{burr96,muel97,kota07}, we first derive useful formulae of the
GW amplitude for axisymmetric emission of neutrinos.

\begin{figure}[htbp]
    \centering
    \includegraphics[width=.5\linewidth]{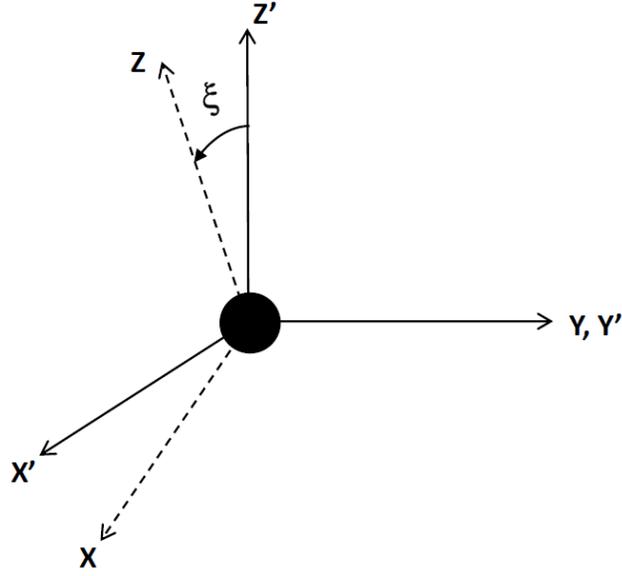}
    \caption{Source coordinate frame $(x',y',z')$ and observer
      coordinate frame $(x,y,z)$. The observer resides at a distant
      point on the $z$-axis.  The viewing angle is denoted by $\xi$,
      which is the angle between $z$ and $z'$ axis. The $z'$-axis
      coincides with the symmetry axis of the source.}
    \label{fig:coordinate}
\end{figure}

First, we introduce the two coordinate frames shown in figure
\ref{fig:coordinate}: the source coordinate frame $(x',y',z')$ and the
observer coordinate frame $(x,y,z)$. In these coordinate frames, the
$z'$-axis is chosen as the symmetry axis of source, while $z$-axis is
set to a line-of-sight direction. Furthermore, the origin of these two
coordinate frames are set to the center of a GRB engine. Thus, the
viewing angle denoted by $\xi$ is given by the angle between the
$z$-axis and the $z'$-axis. For simplicity, we assume that the
$z$-axis lies in $(x',z')$-plane. In this case, the two polarization
states of GWs satisfying the transverse-traceless conditions become
$h_+^\mathrm{TT}\equiv h_{xx}^\mathrm{TT}=-h_{yy}^\mathrm{TT}$ and
$h_{\times}^\mathrm{TT}\equiv h_{xy}^\mathrm{TT}=h_{yx}^\mathrm{TT}$
in the observer coordinate. The geometrical setup shown in Figure
\ref{fig:coordinate} yields the following relationship between the two
polar coordinate frames $(\theta,\phi)$ and $(\theta',\phi')$:
\begin{eqnarray}
    \label{eq:coordinate}
    \sin\theta'\cos\phi'&=&\sin\theta\cos\phi\cos\xi+\cos\theta\sin\xi,\nonumber\\
    \sin\theta'\sin\phi'&=&\sin\theta\sin\phi,\nonumber\\
    \cos\theta'&=&-\sin\theta\cos\phi\sin\xi+\cos\theta\cos\xi.
\end{eqnarray}
With these coordinates, one can obtain the GW amplitude as \citep{muel97}
\begin{eqnarray}
\label{eq:h+}
h_+^\mathrm{TT}(t,\xi)=\frac{2G}{c^4 D}\int^{t-D/c}_{-\infty}dt'\int_{4\pi}d\Omega' 
(1+\cos\theta)\cos(2\phi)\frac{dL_\nu({\bf \Omega}',t')}{d\Omega'},
\end{eqnarray}
where $G$ is the gravitational constant, $c$ is the speed of light,
$D$ is the distance between the observer and the source, and
$dL_\nu/d\Omega'$ represents the direction-dependent neutrino
luminosity per unit of solid angle in the direction of ${\bf\Omega}'$.
Note that the counter part of the amplitude, $h_\times^\mathrm{TT}$,
is obtained by replacing $\cos(2\phi)$ by $\sin(2\phi)$ in Eq.
(\ref{eq:h+}), which vanishes when $dL_\nu/d\Omega'$ is axially
symmetric.

In the above equation, $\theta$ and $\phi$ are required to be
expressed in terms of the angles $\theta'$ and $\phi'$ with respect to
the source coordinate variables, and the viewing angle, $\xi$.  Using
Eq. (\ref{eq:coordinate}), one can obtain the amplitude as following,
\begin{eqnarray}
h_+(t,\xi)=\frac{2G}{c^4 D}\int^{t-D/c}_{-\infty}dt' \int_{4\pi} d\Omega' 
\Psi(\theta',\phi',\xi)\frac{dL_\nu({\bf \Omega}',t')}{d\Omega'},
\end{eqnarray}
where $\Psi(\theta',\phi',\xi)$ denotes the angle dependent factor,
\begin{widetext}
\begin{eqnarray}
    \Psi(\theta',\phi',\xi)=(1+\cos\theta' \cos\xi+\sin\theta' \cos\phi' \sin\xi)
    \frac{(\sin\theta' \cos\phi' \cos\xi-\cos\theta' \sin\xi)^2-\sin^2\theta' \sin^2\phi'}
    {(\sin\theta' \cos\phi' \cos\xi-\cos\theta' \sin\xi)^2+\sin^2\theta' \sin^2\phi'}.
\end{eqnarray}
\end{widetext}
This is a generalization of the Eqs. (26) and (27) of M\"uller \&
Janka \citep{muel97}, who derived the formula for the cases $\xi=0$
and $\pi/2$, respectively.  By integration with respect to the
azimuthal angle ($\phi'$) assuming that $d L_\nu({\bf
  \Omega}',t')/d\Omega'$ is axisymmetric, we obtain
\begin{eqnarray}
h_+(t,\xi)=\frac{2G}{c^4 D}\int^{t-D/c}_{-\infty}dt' \int^{\pi}_{0}\sin\theta' 
d\theta' ~\Phi(\theta',\xi)\frac{dL_\nu}{d\theta'}(\theta',t'), \label{h+}
\end{eqnarray}
where
\begin{widetext}
\begin{eqnarray}
    \Phi(\theta',\xi)=\left\{
    \begin{array}{cc}
        -2\pi\left[1+\cos\theta'\left(2+\cos\xi\right)\right]\tan^2\left(\displaystyle\frac{\xi}{2}\right) & (\mathrm{for}~~\theta\ge\xi)\\
        -2\pi\left[1+\cos\theta'\left(-2+\cos\xi\right)\right]\cot^2\left(\displaystyle\frac{\xi}{2}\right) & (\mathrm{for}~~\theta<\xi)
    \end{array}
    \right. .
\end{eqnarray}
\end{widetext}
When $\xi=\pi/2$, the above equations coincide with Eq. (8) of Kotake
et al. \citep{kota07}. With these equations, we estimate the amplitude
of gravitational waves.  For practical evaluations, we need to know $d
L_{\nu}({\bf \Omega}, t)/d \Omega$ which should be determined by the
nature of the central engine or produced jets.  In the following
subsections, we regard the shape of NDAF as a geometrically thin disk
or an oblate spheroid, by which we obtain the angular dependence of $d
L_{\nu}({\bf \Omega}, t)/d \Omega$ explicitly.

\subsection{A geometrically thin disk model}
We first consider a geometrically infinitely thin disk as a simple but
useful case in order to model NDAF. In this case, the angular
dependence of the neutrino emission is written as $dL_\nu/d\Omega
\propto |\cos\theta|$ if we assume that the emission of neutrinos is
isotropic at the disk surface. We can write the neutrino luminosity
per solid angle as $dL_\nu/d\theta'=(L_\nu/2
\pi)\left|\cos\theta'\right|$, where $L_\nu$ is the {\it total}
neutrino luminosity.

At first, we consider the case of an observer located in the
equatorial plane for simplicity, which means $\xi=\pi/2$. In this case
Eq.(\ref{h+}) is written as
\begin{eqnarray}
h_+(t)&=&\frac{2G}{c^4 D}\int^{t-D/c}_{-\infty}dt' \int^{\pi}_{0}d\theta' 
\Phi(\theta')\frac{dL_\nu}{d\Omega'}(\theta',t')\nonumber\\
    &=& \frac{2G}{c^4 D}\int^{t-D/c}_{-\infty}dt' L_\nu(t')\int^{\pi}_{0} 
d\theta'(-1+2\left|\cos\theta'\right|)\sin\theta'\left|\cos\theta'\right|\nonumber\\
    &=&\frac{2G}{c^4 D}\left(\frac{1}{3}\right)\int^{t-D/c}_{-\infty} L_\nu(t') dt'. 
\label{eq:amplitude}
\end{eqnarray}
It is worthwhile to note that the anisotropy parameter $\alpha$
defined by M\"uller and Janka \citep{muel97} (see Eq.(29) in their
paper) is constant and equal to $1/3$ in this case.  Since the
duration of neutrino emission is finite, the GW amplitude converges to
a non-vanishing value\footnote{This is called {\it burst with memory}
  \citep{brag87}. This nature can be directly seen in the time
  integral in Eq. (\ref{eq:amplitude}). Largely-aspherical mass
  ejection and magnetic fields can lead to a similar GW memory
  \citep{ober06,shib06}. The detectability of such GW bursts with
  memory was discussed in \citep{thor92}.}, which we denote as
$h_{\infty}$. With characteristic values, we can write $h_{\infty}$ as
\begin{eqnarray}
h_{\infty}\sim 1.8\times10^{-21}\left(\frac{10\mathrm{Mpc}}{D}\right)\left(\frac{E_\nu}{10^{54}\mathrm{ergs}}\right) \label{eq:h0},
\end{eqnarray}
where $E_\nu\equiv\int^\infty_{-\infty} L_\nu(t')dt'$ is the total
energy emitted by neutrinos.  In this work, we employ $E_{\nu} =
10^{54}$ ergs for calculations of the GW amplitude. This is because
the required energy for a GRB jet is $E_j \sim 10^{52}$
ergs \footnote{The geometrically corrected gamma-ray energy is smaller
  by one order of magnitude, $E_{\gamma} \sim {10}^{51}$ ergs. But
  note that only the fraction of the jet energy is converted into the
  radiation energy. For example, in the internal shock model, a
  fraction of the jet kinetic energy $\epsilon_{\rm int} \lesssim 0.6$
  can be converted into the internal energy, and a fraction of the
  internal energy is used for acceleration of nonthermal electrons
  (and amplification of the magnetic field). Such large jet energy
  will also be required in some more specific scenarios. For instance,
  the hypothesis that GRBs are the sources of ultra-high-energy cosmic
  rays is typically require $E_{j} \gtrsim {10}^{52}$ ergs.}  and the
efficiency of energy conversion is expected to be $\sim 1$\% for the
neutrino-annihilation mechanism \citep{poph99}.  In addition, the
energy released by matter at innermost stable circular orbit (ISCO) is
about 40\% of its rest-mass energy for an extremely rotating Kerr BH.
This corresponds to $\sim 10^{54}$ ergs for accretion of $1~M_\odot$.
This value is also consistent with the observational evidence of
GRB-HNe association because the explosion energy of HNe is $\sim10$
times larger ($E_{\rm kin} \sim {10}^{52}$ ergs) than that of ordinary
core-collapse SNe, which emit $E_\nu \sim 3\times 10^{53}$ ergs by
neutrinos.  \textit{If} the explosion mechanism of HNe is similar to
that of ordinary SNe, in which the explosion energy is provided by
absorption and scattering of neutrinos, the total energy emitted by
neutrinos must be larger than that of ordinary SNe, and we may expect
$E_{\nu} \sim {10}^{54}$ ergs.

It is useful to see dependence of the GW amplitude on the viewing
angle, $\xi$.  In the thin disk model, Eq.(\ref{eq:h+}) can be
integrated analytically, and we have
\begin{eqnarray}
h_{\infty} (\xi)=\frac{2G}{c^4 D}\frac{1+2\cos \xi}{3}\tan^2\left(\frac{\xi}{2}\right)~E_\nu.
\label{eq:viewangle}
\end{eqnarray}
It is obvious that the GW amplitude is the largest when the observer
is located in the equatorial plane, whereas the GW vanishes when the
observer is located at the pole. This implies important implication,
i.e., anti-correlation between GW signals and prompt gamma-ray
photons.  Prompt photons themselves are highly beamed ($\sim
1/\Gamma$) since they are emitted from ultrarelativistic jets moving
with $\Gamma \sim {10}^{2.5}$.  Therefore, it may not be easy to
detect prompt photons when we can expect the strong GW emission.
Furthermore, there may be the possible population of failed GRBs whose
jets do not generate prompt photons since, e.g., jets are choked in
the stellar envelope. As discussed in the next section, GW signals
would still be useful as a probe of the central engine even in such
cases.

\subsection{An oblate spheroid}
\begin{figure}[htbp]
    \centering
    \includegraphics[width=.5\linewidth]{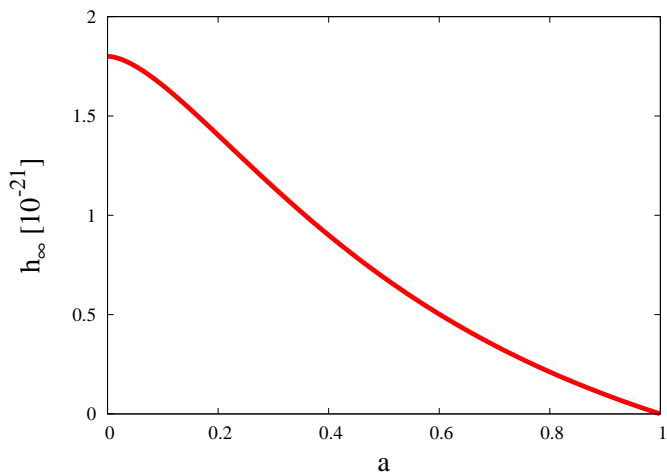}
    \caption{The GW amplitude from NDAF using an oblate spheroid model
      as a function of the ratio between the length of the major and
      the minor axes, $a$. We employ $D=10$ Mpc and $E_\nu=10^{54}$
      ergs in this plot. For spherical emission $(a=1)$, no GW is are
      emitted by neutrinos. The case of the disk height being similar
      to the disk radius corresponds to $a\sim 0.5$ that leads a
      suppression of GW by a factor of 3 compared to the case $a=0$.}
    \label{fig:obl}
\end{figure}

We now consider an oblate spheroid in order to mimic the thickness of
NDAF. In this case, the neutrino luminosity per solid angle is
$dL_\nu/d\Omega\propto\sqrt{a^2+(1-a^2)\cos^2\theta}$, where $a~(\le
1)$ is the ratio between the length of the major and the minor
axes. The gravitational wave amplitude for an observer in the
equatorial plane ($\xi=\pi/2$) can be written as
\begin{widetext}
\begin{eqnarray}
    &&\left.h_{\infty} (a)\right|_{\xi=\pi/2}=\frac{2G}{c^4 D}
     \frac{\sqrt{1-a^2}(1+a+4a^2)-3a^2(1+a)F(a)}{3(1+a)[\sqrt{1-a^2}+a^2 F(a)]}
    E_\nu,\\
    &&F(a)=\log\left(\frac{1+\sqrt{1-a^2}}{a}\right).
\end{eqnarray}
\end{widetext}
Fig \ref{fig:obl} shows the GW amplitude from spheroid as a function
of $a$ with $D=10$ Mpc and $E_\nu=10^{54}$ ergs. For spherical
emission $(a=1)$, no GW is are emitted by neutrinos. The case of the
disk height being similar to the disk radius corresponds to $a\sim
0.5$ that leads a suppression of GW by a factor of 3 compared to the
case $a=0$.

\begin{figure}[htbp]
    \centering
    \includegraphics[width=.5\linewidth]{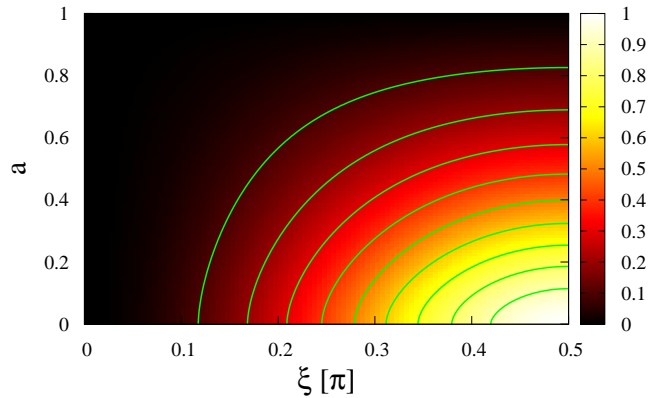}
    \caption{The GW amplitude from NDAF using an oblate spheroid model
      as a function of the ratio between the length of the major and
      the minor axes, $a$, and the viewing angle, $\xi$, normalized to
      its value for $a=0$ and $\xi=\pi/2$.}
    \label{fig:2d}
\end{figure}

In Fig. \ref{fig:2d}, we show $h_{\infty}(\xi,a)$ normalized to its
value for $\xi=\pi/2$ and $a=0$, which is calculated numerically. The
amplitude for $\xi=\pi/4$ and $a=0.5$ is $\sim 0.18
h_{\infty}(\pi/2,0)$.  It should be noted that the suppression is not
so large even if we include both the viewing angle and thickness of
the disk, as long as we consider $a \lesssim 0.5$ and $\xi \gtrsim
\pi/4$, respectively. It should be noted that the amplitude of GW is
anti-correlated with neutrino. The {\it apparent} luminosity of
neutrino shows the largest value for the observer at pole. When we
consider the geometrically thin disk that was discussed in previous
subsection, neutrinos can not be observed for the observer at the
equatorial plane but GW is observable. As for the oblate spheroid,
neutrinos are observable from any direction. The detectability of
neutrinos will be discussed in the next section.

\section{The gravitational wave spectrum and application to bursts} \label{sec:form}
In this section, we first obtain useful formulae of the GW spectrum
from NDAF using the expressions derived in the previous section.  We
now consider axisymmetric sources, so that the only non-vanishing
component is $h_+$. As a result, the local energy flux of GWs can be
written as
\begin{equation}
\frac{dE_\mathrm{GW}}{dA dt}=\frac{c^3}{16 \pi G}\left|\frac{d}{dt}h_+(t)\right|^2,
\end{equation}
where $dA=D^2 d\Omega$ is the surface element. This is the general
expression for the GW flux.

We have calculated the angular dependence of $d L_{\nu}({\bf \Omega},
t)/d \Omega$ in the previous subsections. Then, we can obtain GW
spectra for NDAF mimicked by an infinitely thin disk or an oblate
spheroid for given $L_{\nu}(t)$.  As an example, let us obtain the
useful expression for the case of a geometrically thin disk.
(Numerical calculations are required for the case of an oblate
spheroid.)  Let us insert Eq. (\ref{eq:amplitude}) taking into account
the angular dependence by Eq. (\ref{eq:viewangle}). Then, we get
\begin{equation}
\frac{dE_\mathrm{GW}}{dA dt}=\frac{G}{36\pi c^5 D^2}(1+2\cos\xi)^2 \tan^4\left(\frac{\xi}{2}\right) L_\nu(t)^2.
\end{equation}
The total energy flowing through $dA$ between $t=-\infty$ and
$t=+\infty$ is therefore
\begin{equation}
\frac{dE_\mathrm{GW}}{dA}=\frac{G}{36\pi c^5 D^2}(1+2\cos\xi)^2 
\tan^4\left(\frac{\xi}{2}\right)\int ^\infty_{-\infty}dt L_\nu(t)^2.
\label{eq:dEdt}
\end{equation}
Integrating over a sphere surrounding the source, we find the total
energy emitted by GW as
\begin{equation}
E_\mathrm{GW}=\frac{1}{9}\beta\frac{G}{c^5}\int ^\infty_{-\infty}dt L_\nu(t)^2,
\label{eq:EGW}
\end{equation}
where $\beta=\frac{43}{3}-20\ln2\sim0.47039$. This is consistent with
Eq. (31) in M\"uller \& Janka \citep{muel97} since the anisotropy
parameter $\alpha=1/3$ in this case as already noted in \S II
\footnote{It should be noted that both $\alpha$ and $\beta$ depend on
  $dL_\nu/d\Omega$ so that the configuration of neutrino emission is
  required in order to estimate more precise energy of GW. This is
  different from ordinary quadrupole formula, in which the total
  energy can be calculated just by the time derivative of quadrupole
  wave amplitude $A^{E2}_{lm}$ and $A^{B2}_{lm}$. See \S III in
  M\"uller and Janka \citep{muel97} for detail.}.  In order to get a
GW spectrum, we write $L_\nu$ in Eq. (\ref{eq:EGW}) in terms of the
inverse Fourier transform as
\begin{equation}
L_\nu(t)=\int^\infty_{-\infty}\tilde L_\nu(f)e^{-2\pi i ft}df.
\end{equation}
After several algebraic steps, we easily obtain the GW energy spectrum
as
\begin{equation}
\frac{dE_\mathrm{GW}(f)}{df}=\frac{2}{9}\beta\frac{G}{c^5}\left|\tilde L_\nu(f)\right|^2,
\end{equation}
for a geometrically thin disk model for NDAF.

Once we have a GW spectrum $d E_{\rm GW}/d f$ (obtained numerically or
analytically), we can discuss detectability of the GW. For this
purpose, we shall use the characteristic GW strain expressed as
\begin{equation}
h_c (f)=\sqrt{\frac{2}{\pi^2}\frac{G}{c^3}\frac{1}{D^2}\frac{dE_\mathrm{GW}(f)}{df}}
\label{eq:hc}
\end{equation}
for a given frequency $f$ \citep{flan98}. Since we have this
characteristic strain $h_c (f)$ for either model of NDAF (a thin disk
or an oblate spheroid), we can also compute signal-to-noise ratios
(SNRs) obtained from matched filtering in terrestrial/space
gravitational wave experiments.  The optimal (with respect to the
relative orientation of a source and detector) SNR is given by
\begin{equation}
\mathrm{SNR}^2=\int^\infty_0 d(\ln f)\frac{h_c(f)^2}{h_n(f)^2},
\label{eq:snr}
\end{equation}
where $h_n(f)=[5f S_h(f)]^{1/2}$ is the noise amplitude with $S_h(f)$
being the spectral density of the strain noise in the detector at
frequency $f$. The factor five refers to averaging over all
orientations of the source \citep{flan98}. This determination of the
SNR is optimistic, because a complete set of wave templates required
for the matched filtering analysis is not available for burst
events. However, SNRs for the other techniques of analysis (e.g.,
bandpass filtering and noise monitoring in Ref. \citep{flan98}) can be
connected with the SNR obtained from matched filtering within some
factors.  Thus, Eq. (\ref{eq:snr}) may give a little bit optimistic,
but not severely overestimated values.

In the following subsections, we give explicit expressions of
$L_{\nu}(t)$ in order to apply our formulae to bursts.  First, we show
GW spectra regarding a GRB as one single burst caused by NDAF, where
neutrino emission with $L_{\nu} \sim$~constant is assumed to occur
during the duration of the GRB.  However, the realistic time
dependence of $L_{\nu}(t)$ should depend on the nature of the central
engine, which is unknown.  In fact, the observed complex variability
may be one of the clues to the behavior of the central engine. For
example, the internal shock model, which is one of the most widely
believed scenarios for explaining prompt emission, requires that jets
are highly inhomogeneous (i.e., the dispersion of the bulk Lorentz
factor should be large).  The origin of many subshells moving with
different speeds may be or may not be related to the activity of the
central engine.  In the former case, we may expect that neutrino
emission occurs intermittently\footnote{For another example, flares
  have been observed in many GRBs and they are usually attributed to
  the long time activity of the central engine.}. This motivates us to
consider the case of multiple bursts.
     
\subsection{A single burst}
In this subsection, we consider a single burst event. We assume the
following time evolution of the neutrino luminosity
\begin{eqnarray}
L_\nu(t)=\frac{E_\nu}{T}\Theta(t)\Theta(T-t),
\end{eqnarray}
where $T$ is the duration of the central engine of a GRB and $\Theta$
is the Heaviside step function. This expression corresponds to the
case where the neutrino luminosity from NDAF only depends weakly on
time during the burst.  In this case, the characteristic strain
evaluated by Eq. (\ref{eq:hc}) can be written as
\begin{equation}
    h_c(f)=\frac{\sqrt{\beta}h_\infty}{\pi^2 Tf}\left|\sin(\pi T f)\right|,
    \label{eq:spectrum_single}
\end{equation}
where $h_\infty$ is the converged value of the GW amplitude obtained
by Eq.  (\ref{eq:h0}). The spectrum converges to a constant value of
$\sqrt{\beta}h_\infty/\pi$ for $f \ll 1/T$.  This behavior typically
arises from a sudden change of the neutrino luminosity and is called
zero-frequency limit of GW memory \citep{turn78}.

\begin{figure}[htbp]
    \centering
    \includegraphics[width=.5\linewidth]{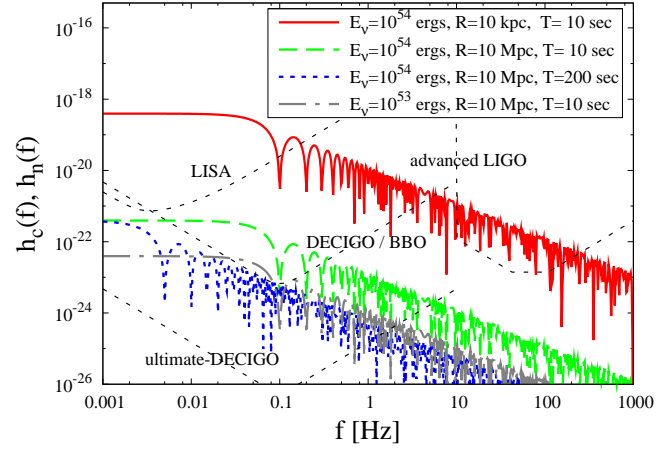}
    \caption{The characteristic amplitude of GWs from a GRB. Thick
      lines represents characteristic amplitudes of GW for typical
      models. The red solid and green dashed lines are for $T=$10 s
      but $D=10$ kpc and 10 Mpc, respectively. The blue dotted line is
      for $D=10$ Mpc and $T$=200 s.  Here, we assume the total energy
      to be $E_\nu=10^{54}$ ergs for red solid, green dashed, and blue
      dotted lines and $E_\nu=10^{53}$ ergs for the gray dot-dashed
      line. }
	\label{fig:spe}
\end{figure}

Figure \ref{fig:spe} shows typical examples of the characteristic
strain $h_c$.  In this figure, the red solid line represents the GW
from GRB, which erupts at the center of Galaxy. The green dashed and
blue dotted lines show the GW from a source at a distance of 10 Mpc
but different durations, $T=10$ and 200 s, respectively.  The gray
dot-dashed line shows the GW from a source with the same parameters as
the green dashed line but $E_\nu=10^{53}$ ergs.  Sensitivity lines for
future planned interferometers, that is, advanced-LIGO, LISA,
DECIGO/BBO, and ultimate-DECIGO, are also shown in this
figure \footnote{The noise curve of advanced LIGO is taken from
  \citep{flan98}. As for LISA and DECIGO/BBO, we adopt
  $S_h(f)=1.22\times
  10^{-51}f^{-4}+2.11\times10^{-41}+1.22\times10^{-37}f^2$\citep{finn00}
  and
  $S_h(f)=4.5\times10^{-51}f^{-4}+4.5\times10^{-45}f^2$\citep{seto01},
  in which $f$ is measured in unit of Hz, respectively.}. One can
roughly read off the SNR from figure \ref{fig:spe}.  It should be
noted that the GW amplitude is larger for the model with
$E_\nu=10^{53}$ ergs and $T=10$ s than one with $E_\nu=10^{54}$ ergs
and $T=200$ s in the range of $f\gtrsim 0.1$ Hz. Therefore, both the
duration and the total energy emitted by neutrinos are important for
the discussion of the detectability.

\begin{figure}[htbp]
    \centering
    \includegraphics[width=.5\linewidth]{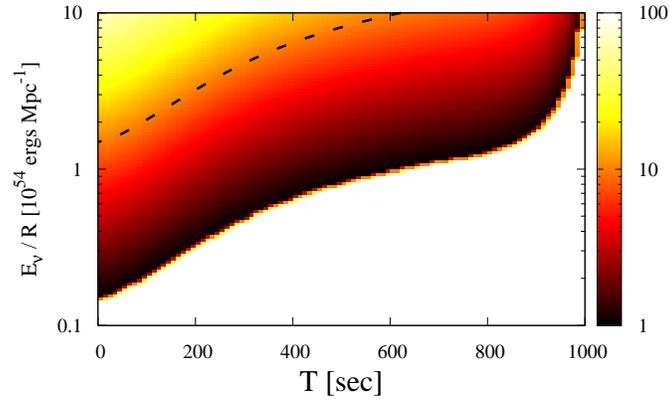}
    \caption{SNRs of GW from NDAF for LISA. The white region in the
      right bottom part of the figure shows SNR$\le 1$. The dashed
      line corresponds to SNR=10.}
    \label{fig:snr}
\end{figure}

\begin{figure}[htbp]
    \centering
    \includegraphics[width=.5\linewidth]{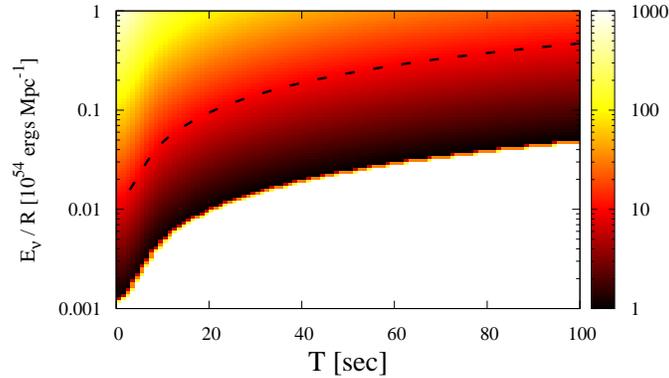}
    \caption{Same as Figure \ref{fig:snr}, but for DECIGO/BBO.}
    \label{fig:snr_decigo}
\end{figure}

More detailed information of the SNR can be found in figure
\ref{fig:snr}, which gives the SNR for LISA. In this figure, the
colored region corresponds to SNR$>1$. The dotted line is for SNR=10,
which is required to detect the GW emission by the burst memory as
discussed in Ref. \citep{thor92}.  As the duration becomes longer, the
SNR of GWs from GRBs gets smaller.  This is because the cutoff
frequency, beyond which GW spectra depend on frequency as $f^{-1}$,
appears at $f\sim T^{-1}$ so that longer lasting events lead to
smaller amplitudes at the high frequency range, whereas amplitudes in
the low frequency range are independent of the duration. From this
figure, one can see that LISA can detect GW signals up to distances of
$D \sim 1$~Mpc. The same analysis shows that GRBs with distances of
$\sim$~100 Mpc are detectable with DECIGO/BBO as long as the duration
$T \lesssim 10$ s (see figure \ref{fig:snr_decigo}).

\subsection{Multiple bursts}
In this subsection, we consider the case of multiple bursts in order
to take into account the intermittent time variability of the central
engine.  We employ the following time evolution for the neutrino
luminosity,
\begin{eqnarray}
L_\nu(t)=\sum_{i=1}^N \frac{E_\nu}{N \delta t}\Theta\left(t-\frac{i}{N}T\right)\Theta\left(\frac{i}{N}T+\delta t-t\right),
\end{eqnarray}
where $N$ is the number of subbursts and $\delta t$ is the duration of
one subburst. We assume $\delta t$ to be constant for any subburst.
The total duration is $T+\delta T\sim T$, since we consider the case
of $\delta T\ll T$ \footnote{Generally speaking, this $\delta t$ may
  not correspond to the observed duration of one subburst. This is
  because the latter can be rather determined by other time scales
  such as the angular spreading time (which becomes important when
  photons are radiated). Note that, if $\delta t \sim T/N$, the
  difference between the two case is diminished.}.  Then, one finds
the characteristic strain as
\begin{eqnarray}
    h_c(f)=\frac{\sqrt{\beta}h_\infty}{\pi^2 N \delta t f}\left|\frac{\sin(\pi\delta t f)\sin(\pi T f)}{\sin(\pi T f/N)}\right|.
    \label{eq:spectrum_multi}
\end{eqnarray}

\begin{figure}[htbp]
    \centering
    \includegraphics[width=.5\linewidth]{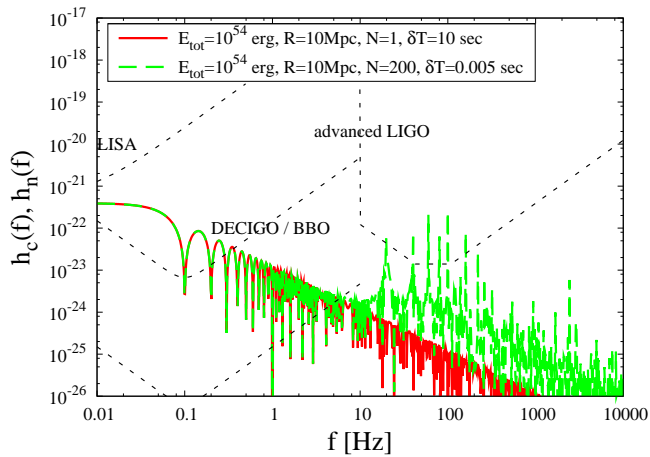}
    \caption{GW spectra of a single burst (red solid line) and
      multiple bursts (green dashed line) from NDAF.}
    \label{fig:multi}
\end{figure}

Figure \ref{fig:multi} shows the GW spectrum of Eqs.
(\ref{eq:spectrum_single}) and (\ref{eq:spectrum_multi}). The red
solid line represents the single burst case, which is the same one as
the green dashed line of figure \ref{fig:spe}. The green dashed line
in this figure shows the case of multiple bursts with $\delta t=0.005$
s and $N=200$. Both of the spectra are shown for $T=10$ s.  We can see
that those two spectra are different in the high-frequency range
because multiple bursts are caused by many bursts with the shorter
time scale.  On the other hand, these spectra coincide with each other
in the low-frequency range because the long-term behavior is
independent from the detail of the burst.  The difference typically
appears only above $f\sim N/T \sim 20$ Hz.  Noting that the most
sensitive frequency of LISA is about 1 mHz, we see that the
detectability by LISA (and DECIGO/BBO) depends mainly $E_\nu$ and $D$,
so that the results of low-frequency GWs described in the previous
subsection are unaltered even if a time variation is considered.  By
contrast, the detectability by advanced-LIGO is affected by
fluctuations with short timescales. The SNR for advanced-LIGO
calculated by Eq. (\ref{eq:snr}) is SNR $\sim 9
(E_\nu/10^{54}\mathrm{erg}) (D/1 \mathrm{Mpc})^{-1}$ for $\delta
t=0.005$ s and $N=200$, which is about 100 times larger than that for
the single burst case. As a result, the GW might be detectable by
advanced-LIGO depending on the time properties of short-term
variations during the burst. This implies that it may be possible not
only to detect GWs by the burst memory, but also to extract important
information about the temporal behavior of the central engine through
GWs.

\subsection{Can we identify NDAF as a gravitational wave source?}
Clearly, it is very important to identify sources emitting GWs in
order to probe the central engine of GRBs and HNe.  However, when we
expect strong GW emission, prompt (electromagnetic) emission may be
off-axis emission which is significantly diminished compared to
on-axis emission.  When the jet angle $\theta_j$ is small enough
compared to the viewing angle $\xi$, the peak flux scales as
${(\varepsilon F_{\varepsilon})}_{\rm max} \propto {[1- (V/c) \cos
    (\xi - \theta_j)]}^{-7/2}$, where V is the velocity of the jet
\cite{yama02}. Then, the peak flux of the off-axis emission is very
roughly estimated as ${(\varepsilon F_{\varepsilon})}_{\rm max} \sim
{10}^{-7}~{\rm ergs}~{\rm cm}^{-2}~{\rm s}^{-1}~L_{\gamma \rm
  max,51.5}^{\rm iso} {(D/10 \rm Mpc)}^{-2}$ for $\xi \sim \pi/4$ and
$\theta_j \sim 0.1$, where $L_{\gamma \rm max,51.5}^{\rm iso}$ is the
luminosity around peak energy (Throughout this subsection we define
$Q_\alpha = Q/10^\alpha$ for a quantity $Q$ in cgs unit).  Although
the details of the spectra and light curves depend on the jet
structure, viewing angle and so on \cite{yama02,gran02}, they may be
detected by the BAT on {\it Swift} or the GBM on {\it Fermi}, if a
bright GRB occurs within 10 Mpc.  However, we may also keep in mind
that many of the discovered nearby GRBs are dimmer than classical
high-luminosity GRBs \cite{lian07,guet07}. For example, if the
luminosity of such a low-luminosity GRB (for on-axis observers) is
dimmer than that of a high-luminosity GRB by more than three orders of
magnitude, and if the jet is well collimated \footnote{It was
  suggested that the outflow of GRB 060218 may not be well collimated
  \cite{sode06b,lian07,guet07}.}, it may be difficult to detect prompt
emission even for very nearby GRBs.

\begin{figure}[htbp]
    \centering
    \includegraphics[width=.5\linewidth]{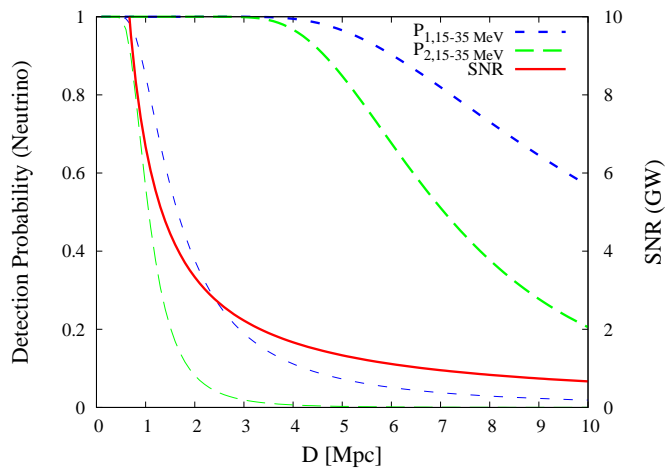}
    \caption{The detection probability of neutrinos and SNR of GWs
      from NDAF. We assume $E_{\nu}={10}^{54}$ ergs.  See text for
      detail.  }
    \label{fig:neu}
\end{figure}

If we can detect both of the GW and prompt emission, we may infer the
distance to the source by follow-up observations \footnote{It seems
  difficult to identify the position of the source only with GW
  observations only, even if we have two or three GW observatories in
  the future.}, and then we can obtain some important information on
the energy budget of the central engine such as $E_{\nu}$ and the
efficiency of energy deposition from $E_{\gamma}$.  Although we may
expect detectable prompt X or gamma rays from such nearby GRBs whose
GW signals are detectable, we can think of a different possibility,
that GWs and neutrinos are expected, but no prompt photons.  In the
collapsar scenario, the jets have to penetrate the progenitor star to
make the prompt emission. But, if they are chocked inside the
progenitor, we do not expect any prompt emission (which is called a
failed GRBs) \cite{mesz01}.  The possible population of such failed
GRBs may not be neglected, as their local rate might be much higher
than the rate of classical high-luminosity GRBs. Even in such cases,
we may still expect detectable GW emission as long as the jets are
launched by the neutrino-annihilation mechanism.  In addition,
coincident neutrino emission from NDAF should be expected, if GWs are
caused by the neutrinos from NDAF.  They may potentially help us to
pin down the position of the bursts, and give us a clue whether GWs
are caused by anisotropic neutrino emission (provided we know the
distance). In figure \ref{fig:neu}, we plot detectabilities of 15-35
MeV neutrinos by the Super-Kamiokande (SK) and Mton detectors as well
as the SNR of GW signals expected by LISA. P$_1$ is the cumulative
Poisson probability to detect at least one neutrino event,
$\sum_{n=1}^\infty (\mu^n e^{-\mu}/n!)$, where $\mu$ is an expected
event number, and P$_2$ is the cumulative Poisson probability to
detect more than one neutrino event, $\sum_{n=2}^\infty (\mu^n
e^{-\mu}/n!)$, in the 15-35 MeV range. Thick lines refer to Mton
neutrino detectors, while thin lines refer to 22.5 kton neutrino
detectors such as SK (see \cite{ando05} for the case of core-collapse
SNe). For simplicity, we have used a Fermi-Dirac distribution for
neutrino spectra with a temperature of 3 MeV, assuming that the
neutrino radiation energy is shared between electron and anti-electron
neutrinos. As we can see, coincident neutrinos can also be
detected. Especially, if we have Mton detectors, we should detect
coincident neutrinos if the GW emission is caused by neutrinos, and
possible detections of scattering events (rather than capturing events
shown in the figure) become useful for positioning (see
\cite{kist08,pagl09} for the discussion of simultaneous observation of
neutrino and GW in the case of core-collapse SNe).

Later (follow-up) observations of photons seem to be crucially
important to determine the distance to the source, as has been already
mentioned.  First, observations in the optical and radio bands may
lead to discovering a HN (e.g., see \cite{sode06a}) associated with
the formation of NDAF making jets. If GRB jets are successful and make
prompt emission, possible detections of the prompt photons would be
helpful for this purpose.  In addition, it may also be possible to
detect afterglow emission (which can be expected not only for
successful GRBs, but also failed GRBs \cite{huan02}).  Whether the
outflow is well collimated or not, it should be decelerated by
colliding with the interstellar medium. When prompt emission is not
observed, afterglows are called \textit{orphan} afterglows. In this
case, the afterglow is also dim in the early phase before the jet
break, during which it may not be so easy to detect its emission.
However, the expected flux after the time when $\Gamma (t_p) \xi= 1$
does not depend on whether afterglows are orphan or not
\cite{huan02,tota02}.  Following the external shock scenario
\cite{mesz06}, we can roughly estimate the peak flux after the peak
time $t_p \sim 8.1 \times {10}^{6}~{\rm s}~[5+2 \ln (10) + 2 \ln
  (\xi/\theta_{j,-1})]{(\xi/\theta_{j,-1})}^{2} \theta_{j,-1}^{8/3}
{(E_{\rm ej, 53}^{\rm iso})}^{1/3} n^{-1/3}$, as ${(\varepsilon
  F_{\varepsilon})}_{\rm max} \sim 6 \times {10}^{-9}~{\rm ergs}~{\rm
  cm}^{-2}~{\rm s}^{-1}~\epsilon_{e,-1} ~E_{\rm ej,53}^{\rm iso}
{(D/\rm Mpc)}^{-2} t_7^{-2}$ for $E_{\rm ej,53}^{\rm iso} \sim
{10}^{53}$ ergs, $\theta_j \sim 0.1$ and $n \sim 1~{\rm cm}^{-3}$,
where $\epsilon_e$ is the fraction of nonthermal electron energy,
$E_{ej}$ is the kinetic energy of the ejecta, and $n$ is the density
of interstellar medium \cite{tota02}, respectively.  Although actual
detectabilities depend on the detail of the afterglow spectra, this
afterglow emission could be detected by groundbased optical telescopes
and X-ray satellites such as \textit{Swift}, future MAXI and EXIST.
Since the spectra and light curves depend on $\theta_j$ and $\xi$,
their detection could potentially give us some information about
relevant parameters such as $\theta_j$ and $\xi$, leading to a crude
estimate of the asphericity parameter $a$ under our NDAF scenario.
Those later observations (i.e., detecting an associated HN component,
prompt and afterglow emissions), combined with GW and neutrino
observations, may help us to reveal the central engine of GRBs and
HNe.

\section{The gravitational wave background}
We are now in a position to discuss the contribution of GWs from NDAF
of GRBs to the background GW radiation. According to Phinney
\citep{phin01}, the sum of the energy densities radiated by a large
number of independent GRBs at each redshift is given by the density
parameter $\Omega_\mathrm{GW}(f)\equiv
\rho_c^{-1}(d\rho_\mathrm{GW}/d\log f)$ as
\begin{eqnarray}
    \Omega_\mathrm{GW}(f)=\frac{1}{\rho_c c^2}\int^{\infty}_0 dz
    \frac{R_\mathrm{GRB}(z)}{1+z}\left|\frac{dt}{dz}\right|f_z\frac{dE_\mathrm{GW}(f_z)}{df},
\end{eqnarray}
where $\rho_c=3H_0^2/(8\pi G)$ is the cosmic critical density,
$R_\mathrm{GRB}(z)$ is the GRB rate per comoving volume, and $f_z
\equiv (1+z)f$. The cosmological model enters with
$|dt/dz|=[(1+z)H(z)]^{-1}$ and, for a flat geometry,
$H(z)=H_0[\Omega_\Lambda+\Omega_M(1+z)^3]$. We adopt the cosmological
parameters $\Omega_M=0.3$, $\Omega_\Lambda=0.7$, and $H_0=100~h_0$
km~s$^{-1}$~Mpc$^{-1}$ with $h_0=0.72$.  For the GRB rate, we employ
the following GRB rate history (GRB3 model in Refs.
\cite{porc01,mura07}) in units of yr$^{-1}$ Gpc$^{-3}$,
\begin{eqnarray}
    R_\mathrm{GRB} &=& R_0 \frac{46 \e^{3.4 z}}{e^{3.8 z}+45}F(z,\Omega_m,\Omega_\Lambda),
\end{eqnarray}
where
$F(z,\Omega_m,\Omega_\Lambda)=\sqrt{\Omega_\Lambda+\Omega_m(1+z)^3}/(1+z)^{3/2}$.
Note that the dependence on the overall GRB rate history is so small
that it is enough to use the above history for our purpose of
demonstration.

\begin{figure}[htbp]
    \centering
    \includegraphics[width=.5\linewidth]{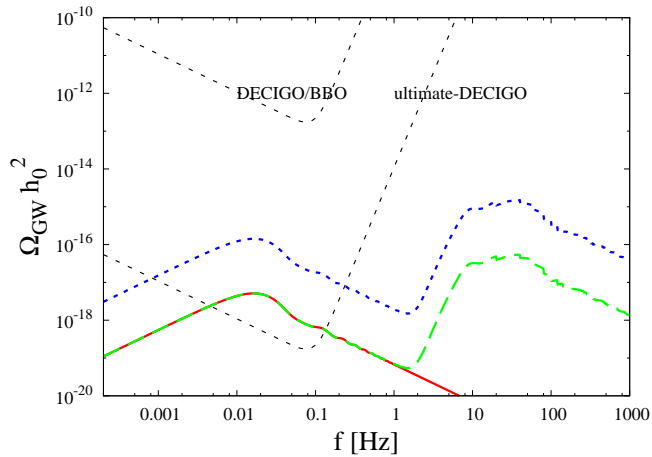}
    \caption{The energy density parameter of gravitational wave
      background (GWB). See text for detail.}
    \label{fig:gwb}
\end{figure}

In figure \ref{fig:gwb}, the calculated $\Omega_\mathrm{GW}$ is
plotted together with the sensitivity curves of future detectors.  We
assume that all GRBs have $T=10$ sec for simplicity.  The red solid
and green dashed lines are obtained for a GRB rate of $R_0=18$ which
is expected for high-luminosity GRBs, while the blue dotted line is
with one for $R_0=500$ which may be expected for low-luminosity GRBs.
The red solid line is obtained with the single-burst spectrum (Eq.
(\ref{eq:spectrum_single})), and the green dashed line is for the
multiple-burst spectrum (Eq. (\ref{eq:spectrum_multi})) so that these
lines differ above 1 Hz. The blue dotted line is for the multiple
burst spectrum. The enhancement of the GW background due to inclusion
of time variation does not change the detectability by space-based
interferometers since the most sensitive frequency is about 0.1 Hz.

Although we have used the inferred rates for successful GRBs, we may
expect a larger contribution to the GW background if there is large
population of failed GRBs.  At present it is difficult to predict the
rate of such choked jets, but we might expect that the rate of failed
GRB is typically smaller than the SN Ibc rate. Then, the possible
contribution to the GW background is at most $\Omega_{\rm GW} h_0^2
\lesssim {10}^{-15}$ at $\sim 0.01-0.1$ Hz (see \cite{naga03,mura07}
for a discussion of the neutrino background from GRBs and HNe).

\section{Summary and Discussions}

In this paper, we have calculated the amplitudes and spectra of GWs
from gamma-ray bursts, which are expected in the neutrino-annihilation
mechanism above an NDAF disk in the collapsar scenario.  We show that
GWs can be detected by LISA and DECIGO/BBO if the source distance is
$D \lesssim$~a few Mpc.  The expected rates of high- and
low-luminosity GRBs occurring at $D< 10$~Mpc, which correspond to the
rates of bursts that can be observed by $\sim 5$~Mton neutrino
detectors \cite{kist08}, are $\sim 2 \times {10}^{-5}~{\rm yr}^{-1}$
and $\sim {10}^{-4}-{10}^{-3}~{\rm yr}^{-1}$, respectively. Hence,
detections of GW signals are possible only when GRBs fortunately occur
in the nearby universe. Nevertheless, the possible GW signals are very
important in the sense that they can probe the mechanism of the GRB
central engine.  Obtained GW spectra are \textit{flat} at low
frequencies, which is the characteristic feature of the GW memory
effect.  In order to detect GWs with LISA or DECIGO/BBO, sufficiently
large values of the anisotropy parameter $\alpha$ and large values of
the released burst energy are typically required.  Since the MHD
mechanism may typically lead to a smaller burst energy compared to the
neutrino-annihilation mechanism (where $E_{\nu} \sim {10}^{53-54}$
ergs is needed), detections of flat GW spectra at low frequencies from
extragalactic GRBs would allow us to infer that the central engine is
governed by the neutrino-annihilation mechanism above an NDAF disk. We
show that coincident neutrinos should be detected if we have Mton
detectors as long as GWs arise from the neutrino emission, which
should also give us important clues about the central engine (e.g.,
estimate of $E_{\nu}$ allows us to make a consistency check for the
burst energy triggering the GW emission).

In order to identify the source emitting GWs, later electromagnetic
observations will be crucial.  We may see prompt emission, afterglow
emission and supernova emission. Although the former is significantly
diminished when the viewing angle is very large, we may be able to
detect it for very nearby bursts. Afterglow emission is also
detectable even if we have orphan afterglows.  Successful observations
of GWs, neutrinos and photons will allow us to diagnose the central
engine of GRBs and to obtain some clues on the jet production
mechanism. In addition, we can expect that HNe are associated with
GRBs (but it may not be necessarily the case). Detections of HNe in
the optical and/or radio band will give us profound information about
the explosion mechanism of HNe.  Strong GW and neutrino emission can
be expected since the neutrino-annihilation mechanism requires that
the radiated neutrino energy is very large ($E_{\nu} \sim
{10}^{53-54}$~ergs). On the other hand, it would be more difficult to
expect detectable GW signals from the MHD mechanism. Hence, if we have
very nearby GRBs, non-detections of both GWs and neutrinos might
suggest the MHD mechanism rather than neutrino-annihilation which
seems to have some caveats (e.g., the very strong dependence on the
disk temperature and inefficiency in production of fireballs).
Whether signals are detected or not, simultaneous observations by GWs
and neutrinos will be important to probe the central engine of GRBs
and HNe, even though the burst rate is not so large.

In this work, we have used a thin disk or an oblate spheroid model to
mimic NDAF. Such a simplified treatment is sufficient for our purpose,
and the characteristic feature in the low-frequency range will not be
altered even if we employ elaborate numerical simulations. But,
although sufficiently realistic calculations have not been done so
far, further studies and developments may allow to evaluate both the
neutrino flux and the GW amplitude more quantitatively. Also, the
effect of neutrino oscillation, which is neglected in this work,
should be included for more quantitative predictions.  In addition, we
have employed the fiducial energy emitted by neutrinos as
$E_\nu=10^{54}$ ergs, which is also unknown. The assumptions employed
for this value are that the energy of a GRB jet is $E_\mathrm{j} \sim
10^{52}$ ergs and the efficiency of energy conversion from neutrinos
to a jet is 1\% for the neutrino-annihilation mechanism.  Since GRBs
have a variety of energetics, $E_\nu$ might be significantly larger or
smaller than this value.  Detections of GWs and neutrinos are useful
since the signal-to-noise ratios of GW show a linear increase with
$E_\nu$. In addition, we might have information about the efficiency
of energy conversion, which is theoretically important, through
obtaining the jet energy from electromagnetic observations.

Although we have mainly considered production of GRB jets with a
duration of $\sim 10-100$ s, required for explaining prompt emission,
there are additional possibilities of the GW emission. Recent
observations by \textit{Swift} have shown that many GRBs have flares
in the afterglow phase, which suggest that the central engine is
active for a longer time than the duration of the prompt emission
\cite{falc07}.  Flares are also likely to be energetic, and the
radiation energy of some flares is even comparable to that of the
prompt emission.  Possibly, they may be produced by slower and less
collimated jets, and their geometrically corrected energy may also be
comparable to that of the prompt emission. Then, we could also expect
GW emission even in the early afterglow phase.

\acknowledgments We would like to thank H.-Th. Janka, K. Kotake, E.
M\"uller, and N. Sago for stimulating discussion.  This study was
supported by the Japan Society for Promotion of Science (JSPS)
Research Fellowships and the Grants-in-Aid for the Scientific Research
from the Ministry of Education, Science and Culture of Japan (No.
S19104006) and for the GCOE Program ``The Next Generation of Physics,
Spun from Universality and Emergence'' from MEXT.

\end{document}